\documentclass[12pt]{article}
\usepackage{amssymb,amsmath,epsfig}

\begin{document}

\title{\bf K-essence Models and Cosmic Acceleration in Generalized Teleparallel Gravity}
\author{M. Sharif \thanks {msharif.math@pu.edu.pk} and Shamaila
Rani \thanks{shamailatoor.math@yahoo.com}\\
Department of Mathematics, University of the Punjab,\\
Quaid-e-Azam Campus, Lahore-54590, Pakistan.}

\date{}

\maketitle
\begin{abstract}
The generalized teleparallel gravity has been suggested to explain
the present cosmic acceleration of the universe. In this paper, we
take spatially homogenous and anisotropic Bianchi type $I$ universe
in the framework of $F(T)$ gravity. The behavior of accelerating
universe is investigated for three purely kinetic k-essence models.
We explore equation of state parameter and deceleration parameter
for these k-essence models. It is found that all these models
exhibit quintessence behavior of the universe.
\end{abstract}

\section{Introduction}

The fact that the universe is expanding at every point in space is a
difficult concept to grasp. Cosmological observations from cosmic
microwave background radiation (CMBR) \cite{1} reveal that most of
the energy in our universe is dark which causes gravitational
repulsion and hence accelerates expansion of the universe. \cite{2}.
The properties of dark energy (DE) can be specified by energy
density $\rho$ and pressure $p$. These two parameters are
responsible for the following three main phases of the universe.

\begin{itemize}
\item The first phase is referred to as radiation dominated era
which occurred just after creation of the universe. At this stage
pressure of the radiation is given by one third of its energy
density.
\item The next is matter dominated era which came into being when universe
was assumed to be $70,000$ years old. Its energy density surpassed
the energy density of the first phase of the universe until a
cosmological constant as a DE candidate was proposed.
\item The third era is DE dominated era. About $5$ billion years ago this phase dominated
in the universe as a whole just after the matter dominated era which
was dropped to very low concentration \cite{3}. It is mentioned here
that the most recent Wilkinson Microwave Anisotropy Probe (WMAP)
observations \cite{2} indicate about $74\%$ of DE in the universe.
\end{itemize}

By the measurements of CMBR, the WMAP satellite indicates that the
universe is very close to the flat and to maintain this flatness,
the mass/energy density of the universe must be equal to a certain
critical density. The total amount of matter in the universe is
estimated only $30\%$ of the critical density. For the remaining
$70\%$ of critical density, an additional form of energy is
required which is termed as DE. There are two proposed forms to
discuss DE: one is the modified theories of gravity and the second
is the scalar field models. In this connection, single scalar
field models attracted many people. K-essence (k stands for
kinetic) model \cite{4} is one of the models which can be
described by a single real scalar field $\phi$ with non-canonical
kinetic term responsible for negative pressure. This may be taken
as the generalization of canonical scalar fields, e.g.,
quintessence \cite{5} (a time varying quantity).

The Lagrangian with a non-canonical kinetic term was proposed to
discuss the early time acceleration named as k-inflation \cite{6}.
Nojiri \cite{7} constructed explicitly k-essence DE model to unify
the late-time acceleration and inflation in early universe.
Matsumoto and Nojiri \cite{8} reconstructed the scalar quintessence
model, tachyon DE model, ghost condensation model as the special
cases of k-essence. Armendariz-Picon et al. \cite{9} discussed
essential features of k-essence and developed some examples. The
solution of these examples lead to two results: one in which cosmic
acceleration continues forever and the other in which acceleration
has finite duration. Bose and Majumdar \cite{10} investigated purely
kinetic k-essence and a particular k-essence model with potential
term. They concluded that such a model could generate basic features
of early inflation and DE observational constraints. Yang and Gao
\cite{11} introduced purely kinetic k-essence by means of
Lagrangian. They plotted evolutions of equation of state (EoS) and
speed of sound for particular cases.

Modified theories also play an important role in explaining DE.
There exist many modified gravity theories which may naturally unify
inflation in early universe and late-time acceleration \cite{12}.
$F(T)$ gravity \cite{13} is the modified form of teleparallel
equivalent of General Relativity \cite{14}. This theory is formed by
using Weitzenb$\ddot{o}$ck connection which has no curvature but
only torsion and possesses second order set of the field equations.

Myrzakulov \cite{15} proposed some new models of k-essence in the
framework of $F(T)$ gravity. Tsyba et al. \cite{16} investigated
purely kinetic k-essence for an explanation of cosmic
acceleration. They concluded that for a particular case of scalar
field, the modified gravity and purely kinetic k-essence become
equivalent. Dent et al. \cite{17} investigated this extended
modified gravity at the background and perturbed level and also
explored this theory for quintessence scenarios. Karami and
Abdolmaleki \cite{18} found that EoS parameter of holographic and
new agegraphic $F(T)$ models always cross the phantom-divide line
whereas entropy-corrected model has to experience some conditions
on parameters of the model. Most of the above mentioned work has
been carried out by using FRW universe for the obvious reasons.

In this paper, we discuss cosmic acceleration by using purely
kinetic k-essence and $F(T)$ gravity with Bianchi type $I$ universe.
The format of the paper is as follows: In the next section, we
present some basics of k-essence model and $F(T)$ gravity theory. In
section \textbf{3}, the field equations are formulated. Section
\textbf{4} is devoted to study some k-essence models and also
discuss cosmic acceleration. In the last section, some concluding
remarks are given.

\section{Preliminaries}

In this section, we provide basic concepts of k-essence as well as
$F(T)$ gravity.

\subsection{K-essence Formalism}

There are some scalar fields with non-canonical kinetic terms in
particle physics. The k-essence models are described by a single
scalar field and a kinetic term. The general k-essence action
\cite{6} is of the form
\begin{equation}\label{1}
S=\int d^{4}x\sqrt{-g}L(\phi,X),
\end{equation}
where $\phi$ is the scalar field, $g=det(g_{\mu\nu})$ and $X$ is the
dimensionless kinetic energy term defined by
\begin{equation}\label{2}
X=\frac{1}{2}g^{\mu\nu}\phi_\mu\phi_\nu,\quad \mu,\nu=0,1,2,3,
\end{equation}
where $\phi_\mu=\frac{\partial\phi}{\partial
x^\mu}=\partial_\mu\phi$. It is mentioned here that a Lagrangian can
be a function of any scalar field $\phi$ and $X$, i.e.,
$L=K(\phi,X)$. Here we consider one of the simplest possible
k-essence model, termed as, purely kinetic k-essence with action
\begin{equation}\label{3}
S=\int d^{4}x\sqrt{-g}L(X),\quad L=K(X).
\end{equation}
The energy-momentum tensor of k-essence is obtained by varying this
action with respect to the metric tensor \cite{6}
\begin{equation}\label{4}
T_{\mu\nu}=K_{X}\phi_{\mu}\phi_{\nu}-Kg_{\mu\nu},\quad
K_X=\frac{dK}{dX}.
\end{equation}
The energy-momentum tensor of perfect fluid is
\begin{equation}\label{5}
T_{\mu\nu}=(\rho_k+p_k)u_\mu u_\nu-p_{k}g_{\mu\nu}.
\end{equation}
Here, $\rho_k,~p_k$ are the energy density and pressure of k-essence
respectively and $u_\mu$ is the four-velocity defined by
\begin{equation}\label{6}
u_\mu=\eta\frac{\phi_\mu}{\sqrt{2X}},
\end{equation}
where $\eta=\pm1$ according to the sign of $\dot{\phi}$ positive or
negative respectively. We assume that $\partial_{\mu}\phi$ is
timelike and smooth on interesting scales \cite{5}. Thus we can
associate the energy-momentum tensors of k-essence and perfect
fluid. For $\mu=0=\nu$ and $\mu=1=\nu$ in Eqs.(\ref{4}) and
(\ref{5}), we obtain the following expressions for k-essence energy
density and pressure respectively
\begin{eqnarray}\label{8}
\rho_{k}=2XK_{X}-K,\quad p_k=K.
\end{eqnarray}
This yields the following k-essence EoS parameter
\begin{equation}\label{10}
\omega_k=\frac{p_k}{\rho_k}=\frac{K}{2XK_{X}-K}.
\end{equation}

\subsection{$F(T)$ Theory of Gravity}

Here, we introduce briefly the teleparallel theory of gravity and
its generalization to $F(T)$ gravity. In teleparallel action, the
torsion scalar $T$ is used as the Lagrangian density while in
modified teleparallel gravity, it is promoted to a function of $T$.
Thus the action for $F(T)$ gravity \cite{18} is
\begin{equation}\label{11}
S=\frac{1}{2\kappa^2}\int d^{4}x[eF(T)+L_m],
\end{equation}
where $e=\sqrt{-g},~\kappa^{2}=8\pi G,~G$ is the gravitational
constant and $F(T)$ is the general differentiable function of $T$.
Also, $L_m$ is the Lagrangian density of matter inside the
universe. The torsion scalar is given as
\begin{equation}\label{12}
T=S_{\rho}~^{\mu\nu}T^{\rho}~_{\mu\nu},
\end{equation}
where $S_{\rho}~^{\mu\nu}$ and torsion tensor $T^{\rho}~_{\mu\nu}$
are defined as follows
\begin{eqnarray}\label{13}
S_{\rho}~^{\mu\nu}&=&\frac{1}{2}(K^{\mu\nu}~_{\rho}
+\delta^{\mu}_{\rho}T^{\theta\nu}~_{\theta}-\delta^{\nu}_{\rho}T^{\theta\mu}~_{\theta}),\\
\label{14}T^{\lambda}~_{\mu\nu}&=&\Gamma^{\lambda}~_{\nu\mu}-
\Gamma^{\lambda}~_{\mu\nu}=h^{\lambda}_{i}
(\partial_{\mu}h^{i}_{\nu}-\partial_{\nu}h^{i}_{\mu}).
\end{eqnarray}
Here $h^{i}_{\mu}$ are tetrad components which form an orthonormal
basis for the tangent space at each point $x^\mu$ of the manifold.
Each vector $h_i$ can be identified by its components $h_\mu^i$
such that $h_i=h^\mu_i \partial_\mu$, where index $i$ runs over
$0,1,2,3$ denote the tangent space while $\mu=0,1,2,3$ represent
the coordinate indices on the manifold. These tetrad related to
the metric tensor $g_{\mu\nu}$ by the following relation
\begin{equation}\label{4*}
g_{\mu\nu}=\eta_{ij}h_{\mu}^{i}h_{\nu}^{j},
\end{equation}
where $\eta_{ij}=diag(1,-1,-1,-1)$ is the Minkowski metric for the
tangent space satisfying the following properties
\begin{equation}\label{11*}
h^{i}_{\mu}h^{\mu}_{j}=\delta^{i}_{j},\quad
h^{i}_{\mu}h^{\nu}_{i}=\delta^{\nu}_{\mu}.
\end{equation}
The contorsion tensor, $K^{\mu\nu}_{\rho}$ is equal to the
difference between Weitzenb$\ddot{o}$ck and Levi-Civita connection
defined by
\begin{equation}\label{15}
K^{\mu\nu}~_{\rho}=-\frac{1}{2}(T^{\mu\nu}~_{\rho}
-T^{\nu\mu}~_{\rho}-T_{\rho}~^{\mu\nu}).
\end{equation}
The variation of Eq.(\ref{11}) with respect to tetrad $h_\mu^i$
leads to the following field equations
\begin{equation}\label{16}
[e^{-1}\partial_{\mu}(eS_{i}~^{\mu\nu})-
h^{\lambda}_{i}T^{\rho}~_{\mu\lambda}S_{\rho}~^{\nu\mu}]F_{T}
+S_{i}~^{\mu\nu}\partial_{\mu}(T)
F_{TT}+\frac{1}{4}h^{\nu}_{i}F=\frac{1}{2}\kappa^{2}h^{\rho}_{i}T^{\nu}_{\rho},
\end{equation}
where $F_{T}=\frac{dF}{dT},~F_{TT}=\frac{d^{2}F}{dT^{2}},~
S_{i}~^{\mu\nu}=h^{\rho}_{i}S_{\rho}~^{\mu\nu}$ with antisymmetric
property. The energy-momentum tensor $T_{\mu\nu}$ is given as
\begin{equation}\label{17}
T^\mu_\nu=diag(\rho_m,-p_m,-p_m,-p_m),
\end{equation}
where $\rho_{m}$ and $p_{m}$ denote the usual density and pressure
of matter inside the universe.

\section{Bianchi $I$ Universe and the Field Equations}

The assumption of isotropy does not predict early epoch of the big
bang as the universe does not maintain its isotropic behavior at
very small scales. In order to get a realistic model which
represents an expanding, homogenous and anisotropic universe, we use
Bianchi type I universe model which is a generalization of FRW
metric. Here we study evolution of the universe in the presence of
DE k-essence models. The line element  of Bianchi type I spacetime
is given by
\begin{equation}\label{18}
ds^{2}=dt^{2}-A^{2}(t)dx^{2}-B^{2}(t)dy^{2}-C^{2}(t)dz^{2},
\end{equation}
where the scale factors $A,~B$ and $C$ are functions of cosmic
time $t$ only. Using Eqs.(\ref{4*}) and (\ref{18}), we obtain the
tetrad components as follows \cite{18a}
\begin{equation}\label{19}
h^{i}_{\mu}=diag(1,A,B,C).
\end{equation}
Using Eqs.(\ref{13}) and (\ref{14}) in (\ref{12}) along with
(\ref{18}), we get
\begin{equation}\label{20}
T=-2\left(\frac{\dot{A}\dot{B}}{AB}+\frac{\dot{B}\dot{C}}{BC}+\frac{\dot{C}\dot{A}}{CA}\right).
\end{equation}
The field equations (\ref{16}) for $i=0=\nu$ and $i=1=\nu$ are given
by
\begin{eqnarray}\label{21}
F-4\left(\frac{\dot{A}\dot{B}}{AB}+\frac{\dot{B}\dot{C}}{BC}
+\frac{\dot{C}\dot{A}}{CA}\right)F_{T}=2\kappa^{2}\rho_{m},
\end{eqnarray}
\begin{eqnarray}\nonumber
&&2\left(\frac{\dot{A}\dot{B}}{AB}+2\frac{\dot{B}\dot{C}}{BC}
+\frac{\dot{C}\dot{A}}{CA}+\frac{\ddot{B}}{B}+\frac{\ddot{C}}{C}\right)F_{T}
-4\left(\frac{\dot{B}}{B}+\frac{\dot{C}}{C}\right)\\\nonumber
&&\times\left[\left(\frac{\ddot{A}}{A}
-\frac{\dot{A^{2}}}{A^{2}}\right)\left(\frac{\dot{B}}{B}+\frac{\dot{C}}{C}\right)
+\left(\frac{\ddot{B}}{B}-\frac{\dot{B^{2}}}{B^{2}}\right)\left(\frac{\dot{C}}{C}
+\frac{\dot{A}}{A}\right)\right.\\\label{22}
&&\left.+\left(\frac{\ddot{C}}{C}-\frac{\dot{C^{2}}}{C^{2}}\right)
\left(\frac{\dot{A}}{A}+\frac{\dot{B}}{B}\right)\right]F_{TT}-F=2\kappa^{2}p_{m}.
\end{eqnarray}
The corresponding conservation equation is
\begin{equation}\label{23}
\dot{\rho_{m}}+\left(\frac{\dot{A}}{A}+\frac{\dot{B}}{B}+\frac{\dot{C}}{C}\right)
(\rho_{m}+p_{m})=0.
\end{equation}
The average scale factor $R$ and the mean Hubble parameter $H$
respectively will become
\begin{eqnarray}\label{24}
R=(ABC)^{1/3},\quad
H=\frac{1}{3}\left(\frac{\dot{A}}{A}+\frac{\dot{B}}{B}+\frac{\dot{C}}{C}\right)
=\frac{\dot{R}}{R}.
\end{eqnarray}
The deceleration parameter $q$ is a dimensionless quantity and is
used to describe the accelerating universe
\begin{equation}\label{q}
q=-1-\frac{\dot{H}}{H^2}.
\end{equation}

Any cosmological universe represents an accelerating, decelerating
or expansion with constant velocity for $-1\le q<0$, $q>0$ and $q=0$
respectively. In terms of Hubble parameter and torsion,
Eqs.(\ref{21})-(\ref{23}) can be simplified as
\begin{eqnarray}\label{26}
2\kappa^{2}\rho_m&=&2TF_T+F,\\\label{27}
2\kappa^2p_m&=&(6\dot{H}-T+2J+2L)F_T+2M\dot{T}F_{TT}-F,\\\label{28}
0&=&\dot{\rho}_m+3H(\rho_m+p_m),
\end{eqnarray}
where $L=\frac{\dot{B}\dot{C}}{BC}-\frac{\ddot{A}}{A},~
M=\frac{\dot{B}}{B}+\frac{\dot{C}}{C}$ and
$J=\frac{\dot{A^{2}}}{A^2}+\frac{\dot{B^{2}}}{B^2}+\frac{\dot{C^{2}}}{C^2}$.
Also, Eq.(\ref{20}) can be written as
\begin{equation}\label{29}
T=-9H^{2}+J.
\end{equation}
The case $F(T)=T$ gives the torsion contribution $\rho_{T},~p_{T}$
in Eqs.(\ref{26}) and (\ref{27}) which reduce to the following forms
\begin{eqnarray}\label{30}
&&\rho_{m}+\rho_{T}=\frac{3T}{2\kappa^{2}},\\\label{31}
&&p_{m}+p_{T}=\frac{1}{\kappa^2}\left(3\dot{H}-T+J+L\right).
\end{eqnarray}
If we take $\rho_T=\rho_k,~p_T=p_k$, then these equations become
\begin{eqnarray}\label{33}
&&\rho_{m}+2XK_{X}-K=\frac{3T}{2\kappa^{2}},\\\label{34}
&&p_{m}+K=\frac{1}{\kappa^2}\left(3\dot{H}-T+J+L\right).
\end{eqnarray}
For the sake of simplicity, we assume that
$\rho_m=0=p_m=\kappa^2-1$. Using these values and inserting the
value of $T$ from Eq.(\ref{29}) in (\ref{33}) and (\ref{34}), it
follows that
\begin{eqnarray}\label{37}
2XK_{X}-K&=&\frac{3}{2}\left(-9H^2+J\right),\\\label{38}K
&=&3\dot{H}+9H^2+L.
\end{eqnarray}
Equation (\ref{2}) implies that $X=\frac{1}{2}\dot{\phi}^2$ whose
corresponding scalar function is given by
\begin{equation}\label{7}
\phi=\int{\sqrt{2X}}dt+constant.
\end{equation}
The continuity equation (\ref{5}) for k-essence models leads to
\begin{equation}\label{39}
\dot{\rho_{k}}+3H(\rho_{k}+p_{k})=0.
\end{equation}
Using Eq.(\ref{8}) in this equation, we have
\begin{equation}\label{40}
(K_X+2XK_{XX})\dot{X}+6HXK_X=0
\end{equation}
which is the kinetic k-essence field equation.

As a special case \cite{15}, we take $\phi=\phi_0+\ln
R^{\pm\sqrt{18}}$, where $\phi_0$ is an arbitrary constant.
Consequently, the kinetic term takes the form
\begin{equation}\label{40+}
X=\frac{1}{2}\dot{\phi}^2=J-T.
\end{equation}
Inserting this value in Eq.(\ref{40}), we get
\begin{equation}\label{40++}
2T(\dot{J}-18H\dot{H})F_{TT}+(\dot{J}-18H\dot{H}+6HT)F_T=0.
\end{equation}
Also, Eq.(\ref{27}) can be simplified as
\begin{equation}\label{28+}
2M(\dot{J}-18H\dot{H})F_{TT}+(6\dot{H}+9H^2+J+2L)F_T-F=0.
\end{equation}
It is interesting to mention here that if we use Eq.(\ref{40++}) in
(\ref{28+}), it shows the equivalence of $F(T)$ gravity and
k-essence for FRW spacetime \cite{15},\cite{16}. However, these
equation do not provide any such relation for Bianchi type $I$
universe.

\section{K-essence Models}

In this section, we aim to discuss k-essence models in the framework
of modified teleparallel gravity. We take three arbitrary k-essence
models and evaluate kinetic term and the corresponding scalar field.
Also, we formulate EoS parameter and the deceleration parameter for
k-essence models. For this purpose, we take the following particular
values of the scale factors by using the power law \cite{20}
\begin{equation}\label{52}
A(t)=(m_{1}s_{1}t)^{1/m_1},\quad B(t)=(m_{2}s_{2}t)^{1/m_2},\quad
C(t)=(m_{3}s_{3}t)^{1/m_3},
\end{equation}
where $m_i,~s_i$ are positive constants and $i=1,2,3$. The EoS
parameter in Eq.(\ref{10}) can be written as follows
\begin{equation}\label{50}
\omega_k=-1+\frac{2XK_X}{2XK_X-K}.
\end{equation}

\subsection{Model I}

Consider the following k-essence model \cite{16}
\begin{equation}\label{41}
K=\sum^{M}_{j=0}\nu_j(t)y^j,\quad M>0,
\end{equation}
where $y=\tanh{t}$. We find the coefficients of $y$ and the kinetic
term of k-essence ($X$) by assuming the following Hubble parameter
\begin{equation}\label{42}
H=\sum_{j=0}^{N}\mu_j(t)y^j,\quad N>0.
\end{equation}
Notice that we take $\mu_j(t)$ and $\nu_j(t)$ as constant
quantities throughout. As a particular example, we take $M=2$ in
Eq.(\ref{41}) and $M=1$ for $H$, it follows that
\begin{eqnarray}\label{43}
K=\nu_0+\nu_{1}y+\nu_{2}y^2,\quad H=\mu_0+\mu_{1}y.
\end{eqnarray}
Inserting this value of $H$ in Eq.(\ref{38}), we obtain
\begin{equation}\label{45}
K=3\mu_{1}(3\mu_{1}-1)y^{2}+18\mu_{0}\mu_{1}y+3\mu_{1}+9\mu_{0}^2+L.
\end{equation}
Using the values of $K$ and $H$ in Eq.(\ref{37}), it follows that
\begin{equation}\label{46}
X=a_{1}e^{\int g_1(t)dt},
\end{equation}
where $a_1$ is an integration constant and
\begin{eqnarray}\nonumber
g_1(t)&=&[3\mu_1+\frac{45}{2}\mu_0^2+L-\frac{3}{2}J-9\mu_0\mu_1y-3\mu_1
(\frac{3}{2}\mu_1+1)y^2]^{-1}\\\nonumber
&\times&2[18\mu_0\mu_1+3\mu_1+9\mu_0^2+\dot{L}+6\mu_1
(3\mu_1-1)y-18\mu_0\mu_1y^2\\\label{47}
&-&6\mu_1(3\mu_1-1)y^3].
\end{eqnarray}
It is mentioned here that the scalar function $\phi$ for the model
(\ref{41}) can be obtained by using Eqs.(\ref{7}) and (\ref{46}).

The k-essence energy density and pressure in Eq.(\ref{8}) take the
following forms
\begin{eqnarray}\label{48}
&&\rho_k=-\frac{27}{2}\mu_1^2y^2-27\mu_0\mu_1y-\frac{27}{2}\mu_1^2+\frac{3}{2}J,\\\label{49}
&&p_k=3\mu_{1}(3\mu_{1}-1)y^{2}+18\mu_{0}\mu_{1}y+3\mu_{1}+9\mu_{0}^2+L.
\end{eqnarray}
Inserting these values in Eq.(\ref{50}), the EoS parameter becomes
\begin{equation}\label{51}
\omega_k=-1-\frac{3\mu_1-\frac{9}{2}\mu_0^2+L+\frac{3}{2}J-9\mu_0\mu_1y-3\mu_1
(\frac{3}{2}\mu_1+1)y^2}{\frac{3}{2}(9\mu_{0}^2-J)+27\mu_0\mu_1y+\frac{27}{2}\mu_1^2y^2}.
\end{equation}
The viability of this model depends on the possible values of the
parameters in Eq.(\ref{52}). The model shows different behavior
initially for unequal $m_{i}$ but indicates the same behavior at
later times for all values of the parameters. Here we discuss the
simple case which leads to the isotropic universe. Using
Eq.(\ref{52}) in this equation and assuming $\mu_i=1=m_i=s_i$, it
follows that
\begin{equation}\label{53}
\omega_k=-1+\frac{1-11/3t^2+6\tanh{t}+5\tanh^2{t}}{9-3/t^2+18\tanh{t}+9\tanh^2{t}}.
\end{equation}
For late time acceleration, i.e., $t\rightarrow \infty$, this yields
\begin{equation}\label{54}
\omega_k\mid_{t\rightarrow\infty}=-\frac{2}{3}
\end{equation}
\begin{figure}
\center\epsfig{file=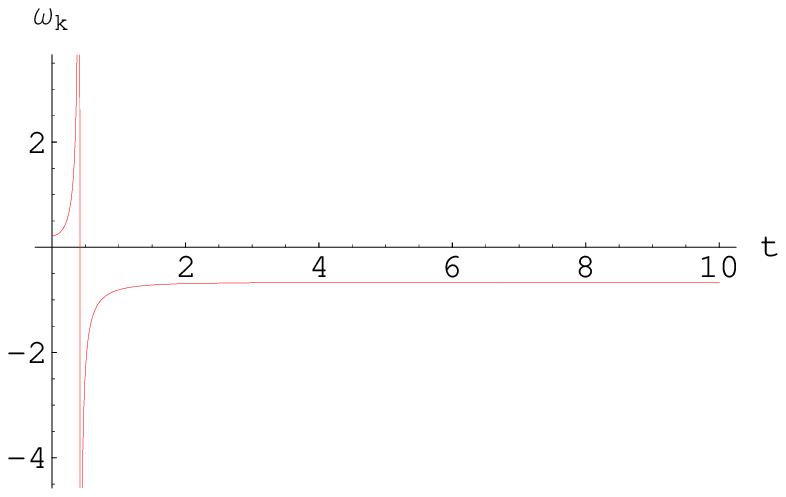, width=0.5\linewidth}\caption{Plot of
$\omega_k$ versus cosmic time $t$ for model I.}
\center\epsfig{file=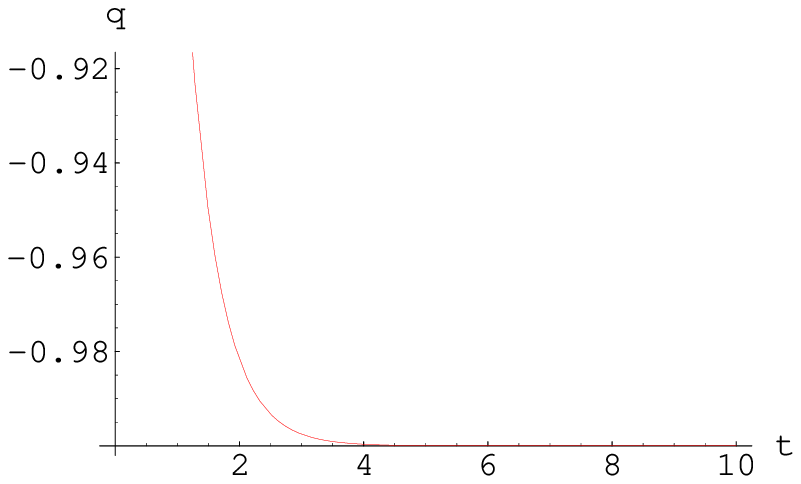, width=0.5\linewidth}\caption{Plot of
$q$ versus cosmic time $t$ for model I.}
\end{figure}
which is shown in \textbf{Figure \ref{1}}. This represents evolution
of k-essence EoS parameter $\omega_k$ as a function of cosmic time.
It is obvious that initially $(t=0),~\omega_k=0.22$ showing that the
universe is lying in a region which contained dust fluid as well as
radiation. As the time elapses up to $t=0.42$, k-essence EoS
parameter confined in the region $-\infty<\omega_k<+\infty$, i.e.,
the universe evolutes from physical matter to DE phase. The EoS
parameter bears a negative increment in its value and becomes
constant at $t=0.48$ towards $t\rightarrow\infty$. The constant
value of EoS parameter is greater than $-1$  but less than $-1/3$
which shows a quintessence era \cite{21}.

The deceleration parameter (\ref{q}) for this model takes the form
\begin{equation}\label{q1}
q=-1+\frac{\mu_1(1-\tanh^2{t})}{\mu_0^2+\mu_1^2\tanh^2{t}+2\mu_0\mu_1\tanh{t}}
\end{equation}
which implies that $q\mid_{t\rightarrow\infty}=-1$ indicating an
accelerating universe. Its graphical representation is shown in
\textbf{Figure \ref{2}} which indicates that initially at
$(t=0),~q=0$ expressing an expanding universe with a constant
velocity. As time passes, its value decreases and converges towards
$-1$. Notice that the graph exhibits an ever expanding universe as
there is not a single positive value of $q$.

\subsection{Model II}

The second k-essence model \cite{15} is given by
\begin{equation}\label{55}
K=\sum^{n}_{j=-m}\nu_{j}(t)e^{jt},
\end{equation}
where $m$ is any positive real number and $n\leq0$ real number. For
the sake of simplicity, we take $m=2,~n=0$ and $\nu_j(t)=\nu_j$ as a
constant. Thus
\begin{equation}\label{56}
K=\nu_{-2}e^{-2t}+\nu_{-1}e^{-t}+\nu_0.
\end{equation}
We also assume $H$ as follows
\begin{equation}\label{57}
H=\mu_{-1}e^{-t}+\mu_0.
\end{equation}
Inserting this value in Eq.(\ref{38}), we get
\begin{equation}\label{58}
K=9\mu_0^2+L-3\mu_{-1}(1-6\mu_0)e^{-t}+9\mu_{-1}^2e^{-2t}.
\end{equation}
Using Eqs.(\ref{57}) and (\ref{58}) in (\ref{37}), we obtain kinetic
term of k-essence model in the form
\begin{equation}\label{59}
X=a_2e^{\int g_2(t)dt}.
\end{equation}
Here $a_2$ is an integration constant and $g_2(t)$ is given by
\begin{equation}\label{60}
g_2(t)=\frac{2[\dot{L}+3\mu_{-1}(1-6\mu_0)e^{-t}-18\mu_{-1}^2e^{-2t}]}{3J/2-9/2+L-3\mu_{-1}
(1+3\mu_{0})e^{-t}-9\mu_{-1}^2e^{-2t}}.
\end{equation}
The scalar function $\phi$ is obtained by inserting Eq.(\ref{59}) in
(\ref{7}).
\begin{figure}
\center\epsfig{file=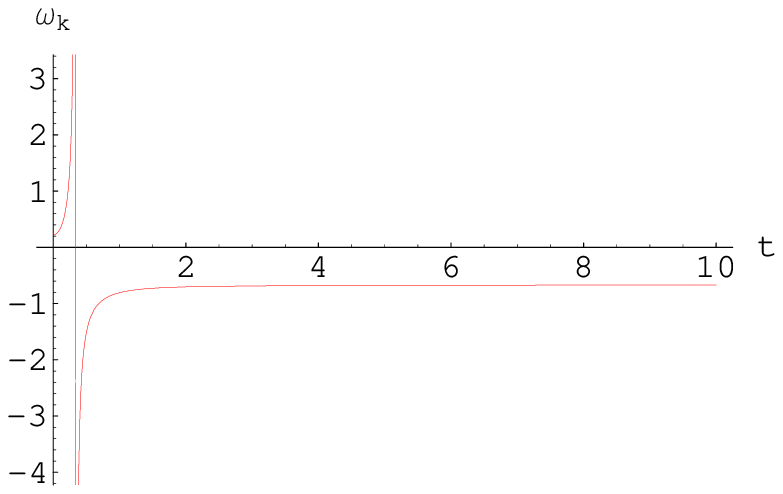, width=0.5\linewidth}\caption{Plot of
$\omega_k$ versus cosmic time for model II.}
\center\epsfig{file=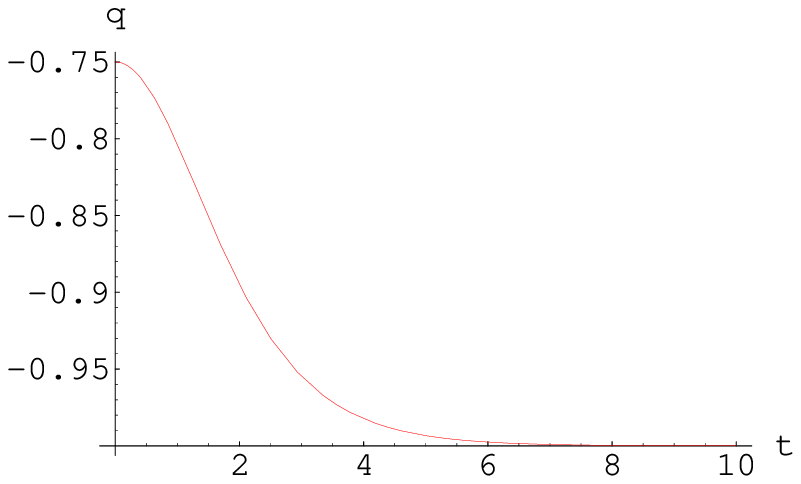, width=0.5\linewidth}\caption{Plot of
$q$ versus cosmic time for model II.}
\end{figure}

The energy density and pressure parameters turn out to be
\begin{eqnarray}\label{61}
\rho_k&=&-\frac{27}{2}\mu_0^2+\frac{3}{2}J-27\mu_{-1}\mu_0e^{-t}-\frac{27}{2}
\mu_{-1}^2e^{-2t},\\\label{62}p_k&=&9\mu_0^2+L-3\mu_{-1}(1-6\mu_0)e^{-t}+9\mu_{-1}^2e^{-2t}.
\end{eqnarray}
Inserting these values in Eq.(\ref{50}), the EoS parameter takes the
form
\begin{equation}\label{63}
\omega_k=-1+\frac{9\mu_0^2-2L-3J+6\mu_{-1}(1+3\mu_{0})e^{-t}+9\mu_{-1}^2e^{-2t}}
{27\mu_0^2-3J+54\mu_{-1}\mu_{0}e^{-t}+27\mu_{-1}e^{-2t}}.
\end{equation}
Taking all the constants equal to $1$ as in isotropic case and
using Eq.(\ref{52}), it follows that
\begin{equation}\label{64}
\omega_k=-1+\frac{9-11/t^2+24e^{-t}+9e^{-2t}}
{27-9/t^2+54e^{-t}+27e^{-2t}}.
\end{equation}
This shows that as $t\rightarrow\infty,~\omega_k>-1$ which
represents quintessence region \cite{21} shown in \textbf{Figure
\ref{3}}. This model has the same behavior as that of the first
model. However, the EoS parameter is directed to DE era at $t=0.34$
and after a short interval of time, it becomes constant, i.e.,
$-0.67$. The corresponding deceleration parameter is
\begin{equation}\label{q2}
q=-1+\frac{\mu_{-1}e^{-t}}{\mu_{-1}^2e^{-2t}+\mu_{0}^2+2\mu_{-1}\mu_{0}e^{-t}},
\end{equation}
which gives $q\mid_{t\rightarrow\infty}=-1$. At $t=0$, the value of
$q$ is negative which shows initially an expanding universe. Its
value decreases and converges to $-1$ as $t\rightarrow\infty$ with
the passage of time as shown in \textbf{Figure \ref{4}}.

\subsection{Model III}

Here we take the following k-essence model \cite{15}
\begin{equation}\label{65}
K=\sum^{n}_{j=-m}\nu_{j}(t)(\ln{t})^j,
\end{equation}
where $m,~n$ and $\nu_j$ are the same as in model II. For
$m=2,~n=0$ and constant $\nu_j$'s, this equation becomes
\begin{equation}\label{66}
K=\nu_{-2}(\ln{t})^{-2}+\nu_{-1}(\ln{t})^{-1}+\nu_0.
\end{equation}
Assuming the Hubble parameter in the form
\begin{equation}\label{67}
H=\mu_{-1}(\ln{t})^{-1}+\mu_0.
\end{equation}
Inserting this value of $H$ in Eq.(\ref{38}), it implies that
\begin{equation}\label{68}
K=L+9\mu_0^2+18\mu_{-1}\mu_0(\ln{t})^{-1}+3\mu_{-1}(3\mu_{-1}-1/t)(\ln{t})^{-2}.
\end{equation}
The kinetic term $X$ from (\ref{37}) and the corresponding scalar
function $\phi$ become
\begin{eqnarray}\label{69}
X=a_3e^{\int g_3(t)dt},\quad \phi=\partial_{t}^{-1}\sqrt{-2X},
\end{eqnarray}
where $g_3(t)$ is found by using (\ref{37})
\begin{eqnarray}\label{71}
&&g_3(t)=2[\dot{L}-6\mu_{-1}(3\mu_{-1}-1/t)t^{-1}(\ln{t})^{-3}+
3\mu_{-1}(1/t-6\mu_{0})t^{-1}(\ln{t})^{-2}]\nonumber\\
&&\times[L+3J/2-9\mu_0^2/2-9\mu_{-1}\mu_{0}
(\ln{t})^{-1}+3\mu_{-1}(-3\mu_{-1}/2-1/t)(\ln{t})^{-2}]^{-1}.\nonumber\\
\end{eqnarray}
\begin{figure} \center\epsfig{file=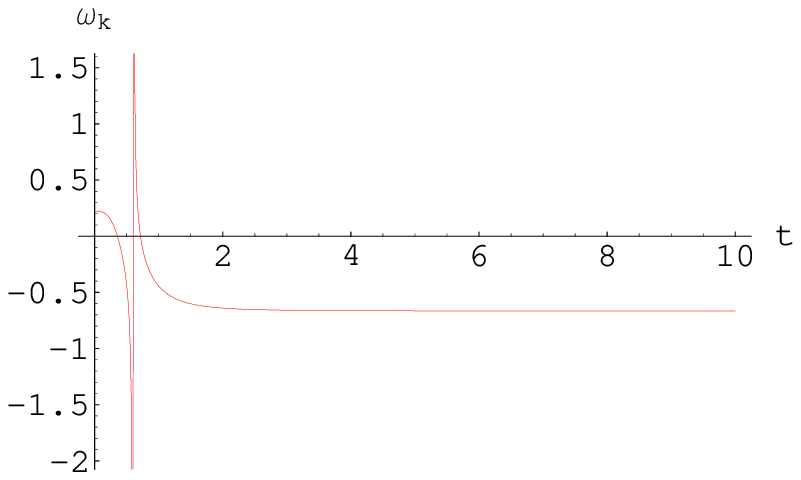,
width=0.5\linewidth}\caption{Plot of $\omega_k$ versus cosmic time
for model III.} \center\epsfig{file=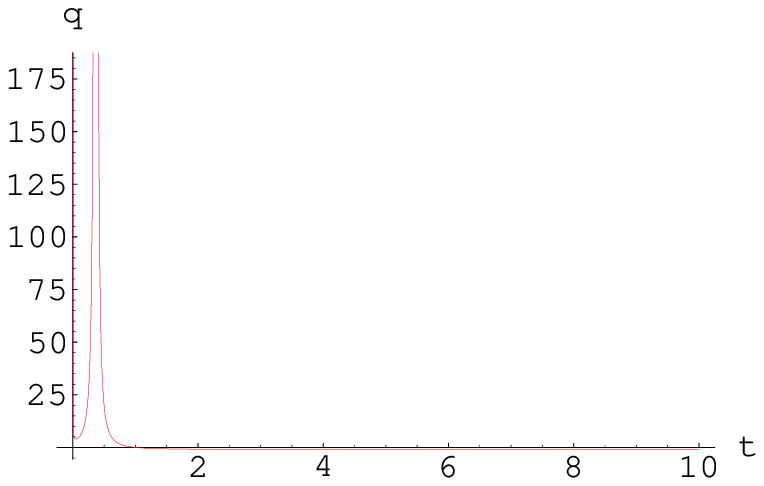,
width=0.5\linewidth}\caption{Plot of $q$ versus cosmic time for
model III.}
\end{figure}

For this model, $\rho_k$ and $p_k$ become
\begin{eqnarray}\label{72}
\rho_k&=&\frac{3}{2}J-\frac{27}{2}\mu_0^2-27\mu_{-1}\mu_{0}(\ln{t})^{-1}-\frac{27}{2}
\mu_{-1}^2(\ln{t})^{-2},\\\label{73}
p_k&=&L+9\mu_0^2+18\mu_0\mu_{-1}(\ln{t})^{-1}+3\mu_{-1}(3\mu_{-1}-\frac{1}{t})(\ln{t})^{-2}.
\end{eqnarray}
The corresponding EoS parameter takes the form
\begin{equation}\label{74}
\omega_k=-1+\frac{2L+3J-9\mu_0^2-18\mu_{-1}\mu_0(\ln{t})^{-1}-6\mu_{-1}(3\mu_
{-1}/2+1/t)(\ln{t})^{-2}}{3J-27\mu_0^2-54\mu_{-1}\mu_0(\ln{t})^{-1}-27\mu_{-1}^2(\ln{t})^{-2}}.
\end{equation}
To expose the present day nature of the universe, we take
$\mu_i=m_i=s_i=1$ and using Eq.(\ref{52}), it follows that
\begin{equation}\label{75}
\omega_k=-1+\frac{3-11/3t^2+6(\ln{t})^{-1}+2(3/2+1/t)(\ln{t})^{-2}}
{9-3/t^2+18(\ln{t})^{-1}+9(\ln{t})^{-2}}.
\end{equation}
Consequently, the deceleration parameter become
\begin{equation}\label{q3}
q=-1+\frac{\mu_{-1}}{t(\ln{t})^2[\mu_{-1}^2(\ln{t})^{-2}+\mu_0^2+2\mu_0\mu_{-1}(\ln{t})^{-1}]}.
\end{equation}

This shows that the condition for accelerating universe is satisfied
as $t\rightarrow\infty$. \textbf{Figure \ref{5}} (\ref{75}) shows
that $\omega_k>-1$ as $t\rightarrow\infty$ and initially the model
represents a universe having both properties of matter and
radiation. After a very short interval, universe is dominated by DE.
It is worthwhile to mention here that at $t=0.63$, the universe
changed its phase from DE era to a physical matter dominated era. As
compared to the DE phase, the universe stayed in matter dominated
era for a short interval of time. Then a decrement in its value is
observed, the EoS parameter for k-essence attains negative values
and finally converges to a constant value resulting quintessence
region \cite{22} of the universe. \textbf{Figure \ref{6}} (\ref{q3})
represents a decelerating universe at early times as the value of
$q$ is positively increasing or decreasing at that stage. After a
long interval of time, the universe intersects the boundary from
matter to DE era and takes $q=-1$. This corresponds to an
accelerating expanding universe.

\section{Summary}

This paper is devoted to study the well-known phenomenon of the
universe expansion in the context of $F(T)$ gravity. For this
purpose, we have taken Bianchi type $I$ universe and have explored
some purely kinetic k-essence models. In physical cosmology and
gravity, the DE problem has been investigated in terms of
cosmological constant, scalar fields, tachyon, Chaplygin gas,
quintessence and modified gravities etc. Here have used $F(T)$
gravity and scalar field term whose combination has resulted a
quintessence phase.

We have evaluated the EoS parameter and the deceleration parameter
for purely kinetic k-essence models. These are shown in terms of
graphs versus cosmic time by taking particular values of the metric
coefficients. These models describe evolution of the universe from
big-bang to present epoch and are summarized as follows
\begin{itemize}
\item For model I, the EoS parameter and the deceleration parameter indicate an
ever accelerating universe with the passage of time. At $t=0.42$,
the universe changes its phase from matter dominated to DE phase
resulting quintessence region.

\item The model II has the same behavior as that of the model I, however,
phase changes at point $t=0.38$. The deceleration parameter
initially represents an expanding universe with constant speed and
after a short interval, it accelerates with a faster speed.

\item For model III, $\omega_k$ and $q$ initially show decelerating
DE dominated universe which reverses to matter dominated
decelerating universe at $t=0.64$. After a slight increment in time,
the universe shows accelerating recent epoch of DE issue lying in
the quintessence region.
\end{itemize}

The analysis of models reveal that the present day universe is
dominated by DE component which can successfully describe
accelerating nature of the universe consistent with the observations
\cite{23}. These new forms of models may give the equivalent
descriptions of DE to discuss the acceleration of the expanding
universe. These models indicate that the present day universe is
dominated by DE component. The viability of these models depend on
the possible values of parameters. Similarly, we can construct the
observational parameters for other new k-essence models induced by
modified gravity theory. Finally, it is worthwhile to mention here
that all the above results turn out to be the generalization of the
already obtained results for the FRW metric \cite{16}. The graphs of
models depend upon the values of parameters accordingly. For Bianchi
type $I$, the k-essence models in the context of generalized
teleparallel gravity mark out a quintessence DE phase.

\end{document}